\documentclass[12pt, a4paper]{article}
\usepackage[top=1in, bottom=1in, left=1.25in, right=1.25in]{geometry}
\usepackage{booktabs}
\usepackage{multirow}
\usepackage{setspace}
\usepackage{natbib}
\usepackage{authblk}
\usepackage{setspace}
\usepackage{amsmath}
\usepackage{amsthm}
\usepackage{amsfonts}
\usepackage{ascmac}
\usepackage{amssymb}
\usepackage{bm}
\usepackage{color}
\usepackage{booktabs}
\usepackage{enumitem}
\usepackage{type1cm} 
\usepackage{mathrsfs}
\usepackage{comment}

\newcommand{\biblist}{\begin{list}{}
{\listparindent 0.0cm \leftmargin 0.50cm \itemindent -0.50 cm
\labelwidth 0 cm \labelsep 0.50 cm
\usecounter{list}}\clubpenalty4000\widowpenalty4000}
\newcommand{\ebiblist}{\end{list}}

\theoremstyle{plain}

\newcommand{\sep}{,\ }

\title{\bf Test for symmetry in $2\times2$ contingency tables with nonignorable nonresponses}
\author[1]{Kouji Tahata}
\author[2]{Yusuke Ii}
\author[3]{Takahiro Nishiyama}

\affil[1,2]{Department of Information Sciences, Tokyo University of Science}
\affil[3]{Department of Business Administration, Senshu University}
\date{Last update: \today}

\begin{document}
\maketitle

\hrulefill
\begin{abstract}
The McNemar test evaluates the hypothesis that two correlated proportion is common in $2\times2$ contingency tables with the same categories.
This study discusses a test for symmetry in $2\times2$ contingency tables with nonignorable nonresponses.
The proposed method is based on Takai and Kano (2008), which discusses a test for independence because a dependency assumption between the two observed outcomes is required to obtain an identification.
Here, we focus on three models and propose a test for symmetry in $2\times2$ contingency tables with nonignorable nonresponses.

\medskip


\medskip

\end{abstract}
\hrulefill 

\newpage
\section{Introduction}
\label{sec.intro}
Incomplete contingency tables (contingency tables with nonresponses) are often encountered in data analysis.
For example, the crime data in Table \ref{tab.K} have been analyzed by many authors \citep[e.g.,][]{K85, S97}.
The variable $Y_1$ indicates whether a victim was affected by a crime six months prior to the visit (1: no, 2: yes), while the variable $Y_2$ indicates whether a crime occurred within six months after the visit.
Additionally, $M_{i}=I(Y_{i}=missing)$ $(i=1,2)$, where $I(\cdot)$ denotes the indicater function.
\begin{table}[h]
    \centering
    \caption{Victimization results from the National Crime Survey from \cite{K85}.\label{tab.K}}
    \begin{tabular}{ccccc} \hline
    \multicolumn{2}{l}{} & \multicolumn{2}{c}{$M_2=0$} & \multicolumn{1}{c}{$M_2=1$} \\ \cline{3-4}
    \multicolumn{2}{l}{} & \multicolumn{1}{c}{$Y_2=1$} & \multicolumn{1}{c}{$Y_2=2$} & \multicolumn{1}{c}{$Y_2= ?$} \\ \hline
    $M_1=0$ & $Y_1=1$& $392$ & $55$ & $33$ \\
     & $Y_1=2$ & $76$ & $38$ & $9$ \\
    $M_1=1$ & $Y_1= ?$ & $31$ & $7$ & $115$ \\ \hline
    \end{tabular}
\end{table}

Previous studies have analyzed incomplete contingency tables with an emphasis on the association between $Y_{1}$ and $Y_{2}$ for the data in Table \ref{tab.K} \citep{K85,S97}.
\cite{M03} described the relationship between the observed variables ($Y_1$, $Y_2$) and the missing indices ($M_1$, $M_2$) in $2\times2$ contingency tables using decomposable models and discussed the identifiability.
\cite{LF96}, \cite{TK08}, and \cite{L18} discussed the test of independence (null association) in $2\times2$ contingency tables with nonresponses.
Additionally, various topics are discussed in biomedical studies \citep{TNG03, TT04, C09, TL17}.
\cite{PKK14} and \cite{KJK20} studied boundary solutions for nonresponse log-linear models, while \cite{NTTT08} proposed the grouped Dirichlet distributions for statistical analysis of incomplete categorical data.

It should be noted that the data in Table \ref{tab.K} have the same classifications.
Such data are unique in that observations tend to concentrate on the main diagonal cells.
For such data, analyses often consider whether the marginal distribution of $Y_1$ is equal to that of $Y_2$, and the issues of symmetry (or homogeneity) rather than independence arise naturally.
In other words, the primary question of interest is whether the percentage of crimes encountered in the first interview is equal to the percentage of crimes encountered in the second interview.
Rejection of the symmetry hypothesis would suggest that the crime risk of the area has changed.
Thus, we are interested in comparing two marginal distributions.

Many fields compare two correlated proportions.
Correlated proportions arise from, for example, longitudinal studies, and such data can be displayed in a $2\times2$ contingency table.
\cite{Mc47} proposed a chi-squared test statistic for the symmetry (or homogeneity) in a $2\times2$ contingency table.
To analyze square $r \times r$ contingency tables, \cite{B48} derived a chi-squared statistic for symmetry, which indicates the symmetry structure of cell probabilities.
\cite{S55} developed a test statistic for the marginal homogeneity to test the structure of homogeneity between two marginal distributions.
The symmetry and the marginal homogeneity are equivalent for $2\times2$ contingency tables.
See \cite{T20} for more details on square contingency table analysis.

Incomplete categorical data for $Y_1$ and $Y_2$ can be classified into a $2\times2$ contingency table with supplemental margins as shown in Table \ref{tab.K}.
Missingness mechanisms are typically classified as missing completely at random (MCAR), missing at random (MAR), and nonignorable (NI) missingness.
See e.g., \cite{R76}, \cite{LR02}, and \cite{FLP}.
For MCAR data, the McNemar test can be employed after throwing out the missing data (see Section \ref{sec.CR}).
However, the McNemar test using only complete cases is not appropriate for MAR data.
Thus, \cite{L09} extended the correlated proportion Bayesian $p$-value approach to $2\times2$ tables with missing data that are MAR.
Additionally, the unconditional test procedures for testing the marginal homogeneity are discussed in \cite{TT04}.
For NI data, \cite{M03} and \cite{TK08} evaluated the identifiability and independence test, respectively.
However, a symmetry test has yet to be developed.

In this paper, we propose a symmetry test for $2\times2$ contingency tables with nonignorable nonresponses.
The proposed method is based on the results given by \cite{TK08}, which discusses a test for independence, because a dependency assumption between two observed outcomes is required to obtain identification.
In particular, we focus on three models and provide a symmetry test for $2\times2$ contingency tables with nonignorable nonresponses.

This paper is organized as follows.
Section \ref{sec.CR} shows the complete case analysis.
Section \ref{sec.Test} describes the proposed method and recommends a symmetry test for  $2\times2$ contingency tables with missing data, which are NI.
Additionally, a numerical method is proposed.
Section \ref{sec.ex} provides a real-data application.
Section \ref{sec.Si} performs simulations to compare the type I error (power) using all data and just complete cases.
Discussion and concluding remarks are provided in Section \ref{sec.Co}.
The Appendix details the numerical algorithms omitted in Section \ref{sec.nu}.

\section{Complete case analysis}
\label{sec.CR}

For complete case analysis, missing data is ignored.
There are three missing mechanisms as described in Section \ref{sec.intro}: MCAR, MAR, and NI missingness.
NI missingness is also referred to not missing at random (NMAR).
See \cite{R76}, \cite{LR02}, and \cite{FLP} for more details of three missing mechanisms.
If the data are MCAR, the McNemar test will have a type I error, which is consistent with the nominal level after throwing out the missing data.

The hypothesis of marginal homogeneity, $H_0:\Pr(Y_1 =1)=\Pr(Y_2 =1)$, is equivalent to the hypothesis of symmetry $H_0:\Pr(Y_1 =1,Y_2=2)=\Pr(Y_1 = 2,Y_2 =1)$.
For complete case analysis, the McNemar statistic with one degree of freedom is given as
\begin{align*}
    z_{0}^{2} = \frac{\left(76-55\right)^{2}}{76+55}=3.37.
\end{align*}
The hypothesis $H_0$ is not rejected at a significance level of 5\%, indicating that the crime rates between the six months before and after the interview are statistically insignificant.
Additionally, if the data in Table \ref{tab.K} are NMAR, the data, including the supplemental margins, should be analyzed.

\section{Methodology and main results}
\label{sec.Test}
Section \ref{sec.Pre} briefly reviews preliminary studies.
Section \ref{sec.TS} describes the goodness-of-fit test of symmetry.
Section \ref{sec.nu} gives the numerical methods related to the proposed test.
\subsection{Preliminary studies}
\label{sec.Pre}
Because this study adopts the notation of \cite{TK08}, we review it briefly.
Let $Y_1$ and $Y_2$ denote the row and column variables, respectively.
$Y_t$ ($t=1,2$) takes values of 1 and 2.
Let $M_1$ and $M_2$ be the missing indicators corresponding to $Y_1$ and $Y_2$, respectively.
A full array of $Y_{1}, Y_{2}, M_{1}$, and $M_{2}$ yields a $2\times2\times2\times2$ contingency table with cell counts $\{n_{kl, ij}\}$ where $i,j=1,2$, $k,l=0,1$, and $\{\pi_{kl,ij}\}$ denote the corresponding cell probabilities.
However, we can observe only a $2 \times 2$ contingency table with supplemental margins (Table \ref{tab.Ob}).
The observed frequencies may be viewed as being derived from the full array.
Namely, the following relations hold
\begin{align*}
    n_{01,i+} &= n_{01,i1}+n_{01,i2} \quad (i=1,2),\\
    n_{10,+j} &= n_{10,1j}+n_{10,2j} \quad (j=1,2),\\
    n_{11,++} &= \sum_{i}\sum_{j} n_{11,ij}.
\end{align*}
Similar relations also hold among the cell probabilities $\pi_{kl,ij}$'s.

\begin{table}[h]
\centering
\caption{Cell frequencies and probabilities with supplemental margins.\label{tab.Ob}}
  \begin{tabular}{ccccc} \hline
    \multicolumn{2}{l}{} & \multicolumn{2}{c}{$M_2=0$} & \multicolumn{1}{c}{$M_2=1$} \\ \cline{3-4}
    \multicolumn{2}{l}{} & \multicolumn{1}{c}{$Y_2=1$} & \multicolumn{1}{c}{$Y_2=2$} & \multicolumn{1}{c}{$Y_2= ?$} \\ \hline
    $M_1=0$ & $Y_1=1$& $n_{00,11}$ & $n_{00,12}$ & $n_{01,1+}$ \\
     & $Y_1=2$ & $n_{00,21}$ & $n_{00,22}$ & $n_{01,2+}$ \\
    $M_1=1$ & $Y_1= ?$ & $n_{10,+1}$ & $n_{10,+2}$ & $n_{11,++}$ \\ \hline \hline
    $M_1=0$ & $Y_1=1$& $\pi_{00,11}$ & $\pi_{00,12}$ & $\pi_{01,1+}$ \\
     & $Y_1=2$ & $\pi_{00,21}$ & $\pi_{00,22}$ & $\pi_{01,2+}$ \\
    $M_1=1$ & $Y_1= ?$ & $\pi_{10,+1}$ & $\pi_{10,+2}$ & $\pi_{11,++}$ \\ \hline
  \end{tabular}
\end{table}

For Table \ref{tab.Ob}, the total sample size $n$ is
\begin{align*}
    n=\sum_{i} \sum_{j} n_{00,ij}+\sum_{i} n_{01,i+}+\sum_{j} n_{10,+j}+n_{11,++},
\end{align*}
and the following equation holds.
\begin{align}
\label{eq1}
    \sum_{i} \sum_{j} \pi_{00,ij} + \sum_{i} \pi_{01,i+} + \sum_{j} \pi_{10,+j} + \pi_{11,++} = 1.
\end{align}
Here, the symbol ``$\bullet$'' represents the corresponding sum as
\begin{align*}
    n_{00,\bullet\bullet} = \sum_{i}\sum_{j}n_{00,ij}, \quad n_{00,i\bullet} = \sum_{j}n_{00,ij} \quad n_{00,\bullet j}=\sum_{i}n_{00,ij}.
\end{align*}
Let
\begin{align*}
   \bm{\pi} &= (\pi_{00,11},\pi_{00,12},\pi_{00,21},\pi_{00,22},\pi_{01,1+},\pi_{01,2+},\pi_{10,+1},\pi_{10,+2},\pi_{11,++}), \\
   \bm{n} &= (n_{00,11},n_{00,12},n_{00,21},n_{00,22},n_{01,1+},n_{01,2+},n_{10,+1},n_{10,+2},n_{11,++}).
\end{align*}
Then, the likelihood based on the observations in Table \ref{tab.Ob} is given as
\begin{align*}
   L(\bm{\pi}|\bm{n})\propto \prod_{i}\prod_{j} \pi_{00,ij}^{n_{00,ij}} \times \prod_{i} \pi_{01,i+}^{n_{01,i+}} \times \prod_{j} \pi_{10,+j}^{n_{10,+j}} \times \pi_{11,++}^{n_{11,++}}.
\end{align*}

Table \ref{tab.TK} shows Models (a)--(c), which indicate the structure of $\{\pi_{kl,ij}\}$.
These models are derived from functional relations and decomposable graphical models.
The variables $(\alpha,\beta_{1},\beta_{2})$, $(\alpha_{1},\alpha_{2},\beta_{1},\beta_{2})$, and $(\alpha_{1},\alpha_{2},\beta,\gamma)$ can be interpreted as odds of (conditional) missing probabilities under Models (a)--(c), respectively.
See \cite{TK08} for more details.

\begin{table}[h]
    \centering
    \caption{Population cell probabilities.\label{tab.TK}}
      \footnotesize
      \begin{tabular}{ccccc} \hline
        \multicolumn{2}{l}{} & \multicolumn{2}{c}{$M_2=0$} & \multicolumn{1}{c}{$M_2=1$} \\ \cline{3-4}
        \multicolumn{2}{l}{Model (a)} & \multicolumn{1}{c}{$Y_2=1$} & \multicolumn{1}{c}{$Y_2=2$} & \multicolumn{1}{c}{$Y_2= ?$} \\ \hline
        $M_1=0$ & $Y_1=1$& $\pi_{00,11}$ & $\pi_{00,12}$ & $\pi_{00,11}\beta_1+\pi_{00,12}\beta_2$ \\
         & $Y_1=2$ & $\pi_{00,21}$ & $\pi_{00,22}$ & $\pi_{00,21}\beta_1+\pi_{00,22}\beta_2$ \\
        $M_1=1$ & $Y_1= ?$ & $(\pi_{00,11}+\pi_{00,21})\alpha$ & $(\pi_{00,12}+\pi_{00,22})\alpha$ & \begin{tabular}{c} 
    $\pi_{00,11}\alpha\beta_1+\pi_{00,12}\alpha\beta_2$ \\ $+\pi_{00,21}\alpha\beta_1+\pi_{00,22}\alpha\beta_2$
    \end{tabular} \\ \hline
    \multicolumn{2}{l}{Model (b)} & \multicolumn{3}{l}{} \\ \hline
        $M_1=0$ & $Y_1=1$& $\pi_{00,11}$ & $\pi_{00,12}$ & $\pi_{00,11}\beta_1+\pi_{00,12}\beta_2$ \\
          & $Y_1=2$ & $\pi_{00,21}$ & $\pi_{00,22}$ & $\pi_{00,21}\beta_1+\pi_{00,22}\beta_2$ \\
        $M_1=1$ & $Y_1= ?$ & $\pi_{00,11}\alpha_1+\pi_{00,21}\alpha_2$ & $\pi_{00,12}\alpha_1+\pi_{00,22}\alpha_2$ & \begin{tabular}{c}
    $\pi_{00,11}\alpha_1\beta_1+\pi_{00,12}\alpha_1\beta_2$ \\ $+\pi_{00,21}\alpha_2\beta_1+\pi_{00,22}\alpha_2\beta_2$
    \end{tabular} \\ \hline
    \multicolumn{2}{l}{Model (c)} & \multicolumn{3}{c}{} \\ \hline
    $M_1=0$ & $Y_1=1$& $\pi_{00,11}$ & $\pi_{00,12}$ & $(\pi_{00,11}+\pi_{00,12})\beta$ \\
         & $Y_1=2$ & $\pi_{00,21}$ & $\pi_{00,22}$ & $(\pi_{00,21}+\pi_{00,22})\beta$ \\
        $M_1=1$ & $Y_1= ?$ & $\pi_{00,11}\alpha_1+\pi_{00,21}\alpha_2$ & $\pi_{00,12}\alpha_1+\pi_{00,22}\alpha_2$ & \begin{tabular}{c}
    $\pi_{00,11}\alpha_1\gamma+\pi_{00,12}\alpha_1\gamma$ \\ $+\pi_{00,21}\alpha_2\gamma+\pi_{00,22}\alpha_2\gamma$
    \end{tabular} \\ \hline
      \end{tabular}
\end{table}

Under each model, the marginal probabilities $\Pr(Y_1=i,Y_2=j)$ are expressed as
\begin{align*}
    \Pr(Y_1=i,Y_2=j)=
    \begin{cases}
        \pi_{00,ij}(1+\alpha)(1+\beta_j) & \mbox{under Model (a)}, \\
        \pi_{00,ij}(1+\alpha_i)(1+\beta_j) & \mbox{under Model (b)}, \\
        \pi_{00,ij}(1+\alpha_i+\beta+\alpha_i \gamma) & \mbox{under Model (c)}. \\
    \end{cases}
\end{align*} 
Therefore, the complete case estimator $n_{00,ij}/n_{00,\bullet\bullet}$ is biased for the cell probability $\Pr(Y_1=i,Y_2=j)$.
When the missing mechanism is not MCAR, analysis based only on complete data may include bias.
That is, we need to estimate the parameters in Table \ref{tab.TK} to obtain a reasonable estimator of $\Pr(Y_1=i,Y_2=j)$.
The number of free parameters $\bm{\pi}$ is $9-1=8$ under the saturated model because (\ref{eq1}) holds.
On the other hand, noting that (\ref{eq1}) holds, Model (a) has $4+1+2-1=6$ free parameters $(\bm{\pi}_{00},\alpha,\bm{\beta})$, Model (b) has $4+2+2-1=7$ free parameters $(\bm{\pi}_{00},\bm{\alpha},\bm{\beta})$, and Model (c) has $4+2+1+1-1=7$ free parameters $(\bm{\pi}_{00},\bm{\alpha},\beta,\gamma)$ where $\bm{\pi}_{00}=(\pi_{00,11},\pi_{00,12},\pi_{00,21},\pi_{00,22})$, $\bm{\alpha}=(\alpha_{1},\alpha_{2})$, and $\bm{\beta}=(\beta_{1},\beta_{2})$.

\subsection{Symmetry test for data with nonignorable nonresponses}
\label{sec.TS}

Let
\begin{align*}
   \bm{m} &= (m_{00,11},m_{00,12},m_{00,21},m_{00,22},m_{01,1+},m_{01,2+},m_{10,+1},m_{10,+2},m_{11,++})
\end{align*}
and $\hat{\bm{m}}$ denote the expected frequency and the corresponding maximum likelihood estimate (MLE) under a hypothesis, respectively.
The likelihood ratio chi-squared statistic of the hypothesis $H_{M}$ is given as
\begin{multline*}
G^{2}(M) = 2 \left( \sum_{i}\sum_{j} n_{00,ij} \log \frac{n_{00,ij}}{\hat{m}_{00,ij}} + \sum_{i} n_{01,i+} \log \frac{n_{01,i+}}{\hat{m}_{01,i+}} \right. \\
\left. + \sum_{j} n_{10,+j} \log \frac{n_{10,+j}}{\hat{m}_{10,+j}} + n_{11,++} \log \frac{n_{11,++}}{\hat{m}_{11,++}} \right)
\end{multline*}
with the corresponding degrees of freedom.
From Section \ref{sec.Pre}, $G^2$ for Model (a) (denoted by $G^{2}(a)$) is asymptotically distributed as a chi-square distribution with two degrees of freedom, and $G^2(b)$ and $G^{2}(c)$ are asymptotically distributed as a chi-square distribution with one degree of freedom each.

The hypothesis of symmetry is 
\begin{align*}
    H_S: \Pr(Y_1=1,Y_2=2)=\Pr(Y_1=2,Y_2=1).
\end{align*}
It should be noted that $\Pr(Y_1=i,Y_2=j)$ is the marginal probability for data with nonignorable nonresponses.
Under Models (a)--(c), the hypothesis $H_S$ is expressed as
\begin{align*}
       H_{S^{a}} &:  \dfrac{\pi_{00,12}}{\pi_{00,21}} = \dfrac{1+\beta_1}{1+\beta_2} \quad \mbox{under Model (a)},\\
       H_{S^{b}} &:  \dfrac{\pi_{00,12}}{\pi_{00,21}} = \dfrac{(1+\alpha_2)(1+\beta_1)}{(1+\alpha_1)(1+\beta_2)} \quad \mbox{under Model (b)},\\
      H_{S^{c}} &:   \dfrac{\pi_{00,12}}{\pi_{00,21}} = \dfrac{1+\alpha_2+\beta+\alpha_2\gamma}{1+\alpha_1+\beta+\alpha_1\gamma} \quad \mbox{under Model (c)}.
\end{align*}
Thus, we need to consider the symmetry test under each model.

The symbol ``$\ast$'' is an element of set $\{a,b,c\}$.
That is, $\ast \in \{a,b,c\}$.
Here, we consider three types of hypotheses: (i) Model ($\ast$) (denoted by $H_{\ast}$), (ii) Model ($\ast$) with symmetry (namely, $H_{S^{\ast}}$), and (iii) symmetry under the assumption that Model ($\ast$) holds (denoted by $H_{S|\ast}$).
Model ($\ast$) with symmetry has one more constraint than Model ($\ast$).
Thus, $G^{2}(S^{a})$ has a chi-square distribution with three degrees of freedom.
Additionally, $G^{2}(S^{b})$ and $G^{2}(S^{c})$ are chi-square distributions with two degrees of freedom.
Model ($\ast$) with symmetry implies Model ($\ast$).
Therefore, a hierarchical relationship exists between the two models.
From this point, we propose the conditional goodness-of-fit test of the hypothesis of symmetry (namely, $H_{S|\ast}$) under the assumption that Model ($\ast$) holds.
That is
\begin{align*}
G^2(S|\ast)=G^2(S^{\ast}) - G^2(\ast).
\end{align*}
This test statistic is asymptotically distributed as a chi-square distribution with one degree of freedom.

As an example, we demonstrate the maximization of the likelihood under the hypothesis $H_{S^{c}}$.
The likelihood function under Model (c) is given as
\begin{multline*}
L(\bm{\pi}_{00},\bm{\alpha},\beta,\gamma|\bm{n}) \propto \prod_{i}\prod_{j} \pi_{00,ij}^{n_{00,ij}} \times \prod_{i} \left( \pi_{00,i1}\beta+\pi_{00,i2}\beta \right)^{n_{01,i+}} \\
\times \prod_{j} \left(\pi_{00,1j}\alpha_1+\pi_{00,2j}\alpha_2 \right)^{n_{10,+j}} \times \left( \gamma\sum_{i}\sum_{j}\pi_{00,ij}\alpha_{i} \right)^{n_{11,++}}.
\end{multline*}
The kernel of the log likelihood function is
\begin{multline}
    \label{eq.ll.c}
    \sum_{i}\sum_{j} n_{00,ij} \log\pi_{00,ij} + \sum_{i} n_{01,i+} \log(\pi_{00,i1}\beta+\pi_{00,i2}\beta) \\
    +\sum_{j} n_{10,+j} \log(\pi_{00,1j}\alpha_1+\pi_{00,2j}\alpha_2)+n_{11,++}\log \left(\gamma\sum_{i}\sum_{j} \pi_{00,ij}\alpha_i \right).
\end{multline}
(\ref{eq.ll.c}) should be maximized under the constraints
\begin{align*}
    \begin{cases}
        \pi_{00,12}(1+\alpha_1+\beta+\alpha_1\gamma)-\pi_{00,21}(1+\alpha_2+\beta+\alpha_2\gamma)=0,\\
        1-\Bigl(\sum_{i}\sum_{j}\pi_{00,ij}+\sum_{i}(\pi_{00,i1}\beta +\pi_{00,i2}\beta) \\
        \qquad \qquad +\sum_{j}(\pi_{00,1j}\alpha_1+\pi_{00,2j}\alpha_2)
        +\gamma\sum_{i}\sum_{j}\pi_{00,ij}\alpha_i\Bigl) =0.
    \end{cases}
\end{align*}
The MLEs are obtained from the numerical optimization.
The numerical method is described in the next section.

\subsection{Proposed numerical method}
\label{sec.nu}

Numerical optimizations are implemented to obtain the actual values of $G^2(S^{c})$ described in Section \ref{sec.TS}.
Typical EM-type algorithms proposed by \cite{D77} are employed.
Our proposed method is based on the algorithm reported in \cite{TK08}, which is similar to the ECM algorithm proposed by \cite{MR93}.
Our algorithm can handle nonlinear equality constraints.
It should be noted that \cite{TK08} implemented the algorithm under some conditions.

Consider the case of $H_{S^{c}}$ as an example.
Let
\begin{align*}
    \bm{Y}_{obs} &= (n_{00,11}, n_{00,12}, n_{00,21}, n_{00,22}, n_{01,1+}, n_{01,2+}, n_{10,+1}, n_{10,+2}, n_{11,++}),&\\
    \bm{Y}_{miss} &= (n_{01,11}, n_{01,12}, n_{01,21}, n_{01,22}, n_{10,11}, n_{10,12},& \\
    &\qquad \qquad \qquad \qquad \qquad \quad n_{10,21}, n_{10,22}, n_{11,11}, n_{11,12}, n_{11,21}, n_{11,22}).
\end{align*}
$\bm{\theta}$ denotes the parameters of interest in the case of the hypothesis $H_{S^{c}}$, $\bm{\theta} = (\bm{\pi}_{00},\bm{\alpha},\beta,\gamma)$.
The algorithm seeks the maximizer of the observed log-likelihood $\ell(\bm{\theta}|\bm{Y}_{obs})$.
For the hypothesis $H_{S^{c}}$, we partition the parameters $\bm{\Theta}=(\bm{\theta},\bm{\lambda})$, including the Lagrange multiplier $\bm{\lambda}=(\lambda_{1},\lambda_{2})$ into $\bm{\Theta}_{1}=\bm{\alpha}$, $\Theta_{2}=\beta$, $\Theta_{3}=\gamma$, and $\bm{\Theta}_{4}=(\bm{\pi}_{00},\bm{\lambda})$.
At the E-step, we define the $Q^*$-function as
\begin{align*}
    Q^*(\bm{\Theta}|\bm{\Theta}^{(t)})=E\left[\ell(\bm{\theta}|\bm{Y}_{obs},\bm{Y}_{miss})-\bm{\lambda} \bm{h}(\bm{\theta})|\bm{Y}_{obs},\bm{\Theta}=\bm{\Theta}^{(t)} \right],
\end{align*}
where $\bm{\Theta}^{(t)}$ is the currently estimated parameter.
Next, we maximize the $Q^*$-function over $\bm{\Theta}_{i}$ with all the other $\bm{\Theta}_{j}$'s, $j\in\{1,2,3,4\}\backslash\{i\}$, fixed at the most recently estimated values at the M-step.
The constraint function $\bm{h}(\bm{\theta})$ under the hypothesis $H_{S^{c}}$ is defined as
\begin{align*}
    \bm{h}(\bm{\theta}) = \left[
    \begin{array}{l}
    (1+\alpha_1+\beta+\alpha_1\gamma)(\pi_{00,11}+\pi_{00,12})\\
    \qquad \qquad +(1+\alpha_2+\beta+\alpha_2\gamma)(\pi_{00,21}+\pi_{00,22})-1\\
    \pi_{00,12}(1+\alpha_1+\beta+\alpha_1\gamma)-\pi_{00,21}(1+\alpha_2+\beta+\alpha_2\gamma)
    \end{array}
    \right].
\end{align*}
The E-step and M-step under the hypothesis $H_{S^{c}}$ are obtained as follows.
\begin{flushleft}
\textbf{E-step}
\end{flushleft}
The $Q^*$-function is given as
\begin{align*}
    Q^*(\bm{\Theta}|\bm{\Theta}^{(t)}) 
    &=\sum_{i}\sum_{j}n_{00,ij}^{(t+1)}\log \pi_{00,ij}+n_{\alpha_{1}}^{(t+1)}\log \alpha_1+n_{\alpha_{2}}^{(t+1)}\log \alpha_2 \\
    & \quad +n_{\beta}^{(t+1)}\log \beta+n_{\gamma}^{(t+1)}\log \gamma,
\end{align*}
where $\{n_{00,ij}^{(t+1)}\}$, $\{n_{\alpha_{k}}^{(t+1)}\}$, $n_{\beta}^{(t+1)}$, and $n_{\gamma}^{(t+1)}$ are defined as
\begin{align*}
    n_{00,ij}^{(t+1)} &= n_{00,ij} + n_{01,i+}\cdot\frac{\pi_{00,ij}^{(t)}}{\pi_{00,i\bullet}^{(t)}}+n_{10,+j}\cdot\frac{\alpha_i^{(t)}\pi_{00,ij}^{(t)}}{\alpha_1^{(t)}\pi_{00,1j}^{(t)}+\alpha_2^{(t)}\pi_{00,2j}^{(t)}} \\
    & \quad +n_{11,++}\cdot\frac{\alpha_i^{(t)}\pi_{00,ij}^{(t)}}{\sum_{s}\sum_{u} \alpha_s^{(t)}\pi_{00,su}^{(t)}} \quad (i=1,2; j=1,2),\\
    n_{\alpha_{k}}^{(t+1)} &= n_{10,+1}\cdot\frac{\alpha_k^{(t)}\pi_{00,k1}^{(t)}}{\alpha_1^{(t)}\pi_{00,11}^{(t)}+\alpha_2^{(t)}\pi_{00,21}^{(t)}}+n_{10,+2}\cdot\frac{\alpha_k^{(t)}\pi_{00,k2}^{(t)}}{\alpha_1^{(t)}\pi_{00,12}^{(t)}+\alpha_2^{(t)}\pi_{00,22}^{(t)}} \\
    & \quad +n_{11,++}\cdot\frac{\alpha_k^{(t)}\pi_{00,k\bullet}^{(t)}}{\sum_{s} \alpha_s^{(t)} \pi_{00,s\bullet}^{(t)}} \quad (k=1,2),\\
    n_{\beta}^{(t+1)} &= n_{01,1+} + n_{01,2+},\\
    n_{\gamma}^{(t+1)} &= n_{11,++}.
\end{align*}
Note that $n_{\beta}^{(t+1)}$ and $n_{\gamma}^{(t+1)}$ are a constant.
\begin{flushleft}
\textbf{M-step}
\end{flushleft}
\begin{description}
    \item[1st step]\begin{align*}
        \alpha_1^{(t+1)} &=\frac{n_{\alpha_{1}}^{(t+1)}}{\left(\gamma^{(t)}+1\right)\left(\lambda_{1}^{(t)}\pi_{00,1\bullet}^{(t)}+\lambda_{2}^{(t)}\pi_{00,12}^{(t)}\right)}, \\ 
        \alpha_2^{(t+1)} &=\frac{n_{\alpha_{2}}^{(t+1)}}{\left(\gamma^{(t)}+1\right)\left(\lambda_{1}^{(t)}\pi_{00,2\bullet}^{(t)}-\lambda_{2}^{(t)}\pi_{00,21}^{(t)}\right)}.
    \end{align*}
    \item[2nd step]\begin{align*}
        \beta^{(t+1)} =\frac{n_{\beta}^{(t+1)}}{\lambda_{1}^{(t)}\pi_{00,\bullet\bullet}^{(t)}+\lambda_{2}^{(t)}\left(\pi_{00,12}^{(t)}-\pi_{00,21}^{(t)}\right)}.
    \end{align*}
    \item[3rd step]\begin{align*}
        \gamma^{(t+1)} =\frac{n_{\gamma}^{(t+1)}}{\lambda_{1}^{(t)}\left(\sum_{s}\alpha_s^{(t+1)} \pi_{00,s\bullet}^{(t)} \right)+\lambda_{2}^{(t)}\left(\alpha_1^{(t+1)} \pi_{00,12}^{(t)}-\alpha_2^{(t+1)} \pi_{00,21}^{(t)}\right)}.
    \end{align*}
    \item[4th step]\begin{align*}
        \pi_{00,11}^{(t+1)} &=\frac{n_{00,11}^{(t+1)}}{\lambda_{1}^{(t)}\left(1+\alpha_{1}^{(t+1)}+\beta^{(t+1)}+\alpha_1^{(t+1)}\gamma^{(t+1)}\right)},\\
        \pi_{00,12}^{(t+1)} &=\frac{n_{00,12}^{(t+1)}+n_{00,21}^{(t+1)}}{2\lambda_{1}^{(t)}\left(1+\alpha_{1}^{(t+1)}+\beta^{(t+1)}+\alpha_1^{(t+1)}\gamma^{(t+1)}\right)},\\
        \pi_{00,21}^{(t+1)} &=\frac{n_{00,12}^{(t+1)}+n_{00,21}^{(t+1)}}{2\lambda_{1}^{(t)}\left(1+\alpha_{2}^{(t+1)}+\beta^{(t+1)}+\alpha_2^{(t+1)}\gamma^{(t+1)}\right)},\\
        \pi_{00,22}^{(t+1)} &=\frac{n_{00,22}^{(t+1)}}{\lambda_{1}^{(t)}\left(1+\alpha_{2}^{(t+1)}+\beta^{(t+1)}+\alpha_2^{(t+1)}\gamma^{(t+1)}\right)},\\
        \lambda_{1}^{(t+1)} &=n_{00,11}^{(t+1)}+n_{00,12}^{(t+1)}+n_{00,21}^{(t+1)}+n_{00,22}^{(t+1)},\\
        \lambda_{2}^{(t+1)} &=\frac{\lambda_{1}^{(t)}\left(n_{00,12}^{(t+1)}-n_{00,21}^{(t+1)}\right)}{n_{00,12}^{(t+1)}+n_{00,21}^{(t+1)}}.
        \end{align*}
\end{description}
Note that $\lambda_{1}^{(t+1)}$ is the sample size $n$.

After this algorithm converges, the modified EM algorithm produces a fixed point of the system of equations, or a solution.
This must be the MLEs.
In a similar manner to hypothesis $H_{S^{c}}$, we can consider the likelihood ratio chi-squared statistic for the hypotheses $H_{S^{a}}$ and $H_{S^{b}}$.
The details are given in the Appendix.

\section{Real-data application}
\label{sec.ex}
Consider the crime data in Table \ref{tab.K} and apply Models (a)--(c).
The statistic $G^2(a)=296.17$ is greater than 5.99, the statistic $G^2(b)=178.32$ is greater than 3.84, and the statistic $G^2(c)=0.03$ is less than 3.84.
Thus, we adopt Model (c) because it is the only one that seems valid among the three candidates for the data in Table \ref{tab.K}.

Next, (\ref{eq.ll.c}) is maximized using the proposed method.
As a result,
\begin{align*}
\hat{\alpha}_{1}=0.083,\quad \hat{\alpha}_{2}=0.010,\quad \hat{\beta}=0.075,\quad \hat{\gamma}=3.026
\end{align*}
are obtained.
Additionally,
\begin{center}
$\bm{\hat{\pi}}_{00}=(0.515,0.078,0.098,0.051)$
\end{center}
is obtained.
The statistic $G^2(S^{c})=1.01$ is less than $5.99$.
Thus, the hypothesis $H_{S^{c}}$ is accepted at the 5\% significance level.
Here, we are interested in determining whether hypothesis $H_{c}$ or hypothesis $H_{S^{c}}$ is preferable for Table \ref{tab.K}.
Therefore, we use the conditional test.
The conditional test of hypothesis $H_{S|c}$ under the assumption that $H_{c}$ holds is expressed as 
\begin{align*}
    G^2(S|c)=G^2(S^{c}) - G^2(c) = 1.01 - 0.03 = 0.98 < 3.84.
\end{align*}
Analysis using all the data, including missing data that cannot be ignored, shows that the hypothesis is accepted at the 5\% significance levels.
In the analysis using all data, it is inferred that there will be no change because the transition of crime damage over six months is not statistically significant.
This is the same result of a complete case analysis in Section \ref{sec.CR}.
\section{Simulation study}
\label{sec.Si}

We conducted a simulation to compare the symmetry tests for complete data and that for data with nonignorable nonresponses.
We assumed that Model (c) holds true and give parameters $(\bm{\pi}_{00},\bm{\alpha},\beta,\gamma)$ where $\bm{\pi}_{00}=(\pi_{00,11},\pi_{00,12},\pi_{00,21},\pi_{00,22})$ and $\bm{\alpha}=(\alpha_1,\alpha_2)$.
Thus, we can obtain the cell probability 
\begin{align*}
\bm{\pi}^{*}=(\pi_{00,11},\pi_{00,12},\pi_{00,21},\pi_{00,22},\pi_{01,1+},\pi_{01,2+},\pi_{10,+1},\pi_{10,+2},\pi_{11,++})
\end{align*}
with supplemental margins, as described in Table \ref{tab.TK}.

Here, we explain how to set the parameters $(\bm{\pi}_{00},\bm{\alpha},\beta,\gamma)$.
Let
\begin{align*}
	p = \Pr(Y_{1}=1,Y_{2}=2|M_{1}=0,M_{2}=0,Y_{1}+Y_{2}=3).
\end{align*}
The conditional probability $p$ can be expressed as
\begin{align*}
	p = \frac{\pi_{00,12}}{\pi_{00,12}+\pi_{00,21}}.
\end{align*}
If $\bm{\pi}_{00}=(\pi_{00,11},\pi_{00,12},\pi_{00,12},\pi_{00,22})$, that is $\pi_{00,12}=\pi_{00,21}$, then $p=0.5$.
It should be noted that the condition $\pi_{00,12}=\pi_{00,21}$ in Model (c) with symmetry is equivalent to the condition $\alpha_{1}=\alpha_{2}$.
In such a case, it is assumed that both symmetry tests using all the data and that using complete data have type I error rates, which are close to the nominal level. 
Additionally, let
\begin{align*}
	p^{*} = \Pr(Y_{1}=1,Y_{2}=2|Y_{1}+Y_{2}=3).
\end{align*}
The conditional probability $p^{*}$ can be expressed as
\begin{align*}
	p^{*} = \frac{\pi_{00,12}(1+\alpha_{1}+\beta+\alpha_{1}\gamma)}{\pi_{00,12}(1+\alpha_{1}+\beta+\alpha_{1}\gamma)+\pi_{00,21}(1+\alpha_{2}+\beta+\alpha_{2}\gamma)}.
\end{align*}
The hypothesis $H_{S^{c}}$ is rewritten using the conditional probability $p^{*}$ as
\begin{align*}
	H_{S^{c}} : p^{*} = 0.5.
\end{align*}
For any specified $p$ and $p^{*}$
\begin{align}
\label{cond}
	\pi_{00,21}=K \pi_{00,12} \quad \mbox{and} \quad \alpha_{2} = \frac{1-p^{*}}{Kp^{*}}\alpha_{1} + \frac{(1-p^{*}-Kp^{*})(1+\beta)}{Kp^{*}(1+\gamma)},
\end{align}
where $K=(1-p)/p$.
Thus, we are interested in scenarios in which the conditional probability $p^{*}$ is 0.5 or differs from 0.5.
To set the parameter $(\bm{\pi}_{00},\bm{\alpha},\beta,\gamma)$, the values $(p,p^{*},\pi_{00,11},\pi_{00,12},\alpha_{1},\beta,\gamma)$ must be specified due to (\ref{eq1}) and (\ref{cond}).

The simulation considers two scenarios and three kinds of sample sizes $n= 250, 500$, and 1000.
The number of simulation replications was 2,000.
We explored the rejection percentages for a 5\% significance level test using a multinomial sample from {\it Multi$(n,\bm{\pi}^{*})$}.

In Scenario 1, the conditional probability $p$ increases from 0.50 to 0.60 in increments of 0.05, and
\begin{align*}
(p^{*},\pi_{00,11},\pi_{00,12},\alpha_{1},\beta,\gamma) = (0.50,0.50,0.10,0.05,0.15,1.20).
\end{align*}
It should be noted that the hypothesis $H_{S^{c}}$ holds.
That is, $p^{*}=0.50$.
Hence, we explored the type 1 error.
Table \ref{sce1} shows the rejection percentage for Scenario 1.
In the cases of $p=0.50$, the rejection percentages of both the McNemar test and the proposed test are close to the nominal level.
On the other hand, the rejection percentages of the McNemar test increase as the value of $p$ increases, whereas those of the proposed test are close to the nominal level.
The proposed test works well.
 
\begin{table}[h]
\caption{Rejection percentages for Scenario 1 \label{sce1}}
\centering
{\footnotesize
\begin{tabular}{cccccc}
\hline
$n$  &  $p$  &  McNemar  &   $S^{c}$  &  $c$  &  $S|c$ (Proposed test)  \\
\hline
250 & 0.50 & 0.0485  & 0.0510  & 0.0595  & 0.0335    \\
 & 0.55 & 0.0955  & 0.0525  & 0.0640  & 0.0360    \\
 & 0.60 & 0.2405  & 0.0535  & 0.0675  & 0.0350    \\
500 & 0.50 & 0.0600  & 0.0575  & 0.0575  & 0.0480    \\
 & 0.55 & 0.1495  & 0.0480  & 0.0565  & 0.0480    \\
 & 0.60 & 0.4480  & 0.0480  & 0.0570  & 0.0400    \\
1000 & 0.50 & 0.0470  & 0.0460  & 0.0500  & 0.0390    \\
 & 0.55 & 0.2795  & 0.0460  & 0.0535  & 0.0400    \\
 & 0.60 & 0.7405  & 0.0550  & 0.0485  & 0.0530    \\
\hline
\multicolumn{6}{l}{Note: $(p,p^{*},\pi_{00,11},\pi_{00,12},\alpha_{1},\beta,\gamma) = (p,0.50,0.50,0.10,0.05,0.15,1.20)$}
\end{tabular}
}
\end{table}

In Scenario 2, the conditional probability $p^{*}$ increases from 0.50 to 0.60 in increments of 0.05, and
\begin{align*}
(p,\pi_{00,11},\pi_{00,12},\alpha_{1},\beta,\gamma) = (0.50,0.30,0.10,0.25,0.10,1.50).
\end{align*}
We explored the type 1 error when $p^{*}=0.50$ and the power when $p^{*}=0.55, 0.60$.
Table \ref{sce2} shows the rejection percentages for Scenario 2.
In the cases of $p^{*}=0.50$, the rejection percentages of both the McNemar test and the proposed test are close to the nominal level.
The proposed test tends to be more conservative than the McNemar test.
Additionally, the rejection percentages of the McNemar test do not increase as the value of $p^{*}$ increases.
Namely, the rejection percentages of the McNemar test are close to the nominal level.
By contrast, those of the proposed test increase as the value of $p^{*}$ increases.
In this case, the proposed test is more powerful than the McNemar test.

\begin{table}[h]
\caption{Rejection percentages for Scenario 2 \label{sce2}}
\centering
{\footnotesize
\begin{tabular}{cccccc}
\hline  
  $n$	&	$p^{*}$	&	McNemar	&	 $S^{c} $	&	$c$	& $S|c$ (Proposed test) 		\\
  \hline
250 & 0.50  & 0.0525  & 0.0460  & 0.0705  & 0.0240    \\
 & 0.55  & 0.0540  & 0.0515  & 0.0615  & 0.0365  \\
 & 0.60  & 0.0560  & 0.1235  & 0.1015  & 0.0955    \\
500 & 0.50  & 0.0455  & 0.0535  & 0.0595  & 0.0350    \\
 & 0.55  & 0.0515  & 0.0820  & 0.0630  & 0.0730    \\
 & 0.60  & 0.0480  & 0.2070  & 0.0970  & 0.2230    \\
1000 & 0.50  & 0.0495  & 0.0475  & 0.0500  & 0.0435    \\
 & 0.55  & 0.0485  & 0.0825  & 0.0510  & 0.0940    \\
 & 0.60  & 0.0430  & 0.3690  & 0.0770  & 0.4590    \\
\hline
\multicolumn{6}{l}{Note: $(p,p^{*},\pi_{00,11},\pi_{00,12},\alpha_{1},\beta,\gamma) = (0.50,p^{*},0.30,0.10,0.25,0.10,1.50)$}
\end{tabular}
}
\end{table}

\section{Conclusion}
\label{sec.Co}

We propose a test of symmetry for data with nonignorable nonresponses.
We focus on the symmetry structure for the three models described in \cite{TK08} and provide  numerical methods for each model to obtain the MLEs of the expected frequencies.
These tests are useful for data with nonignorable nonresponses for two reasons.
Unlike the usual test using complete data with a type I error rate, which gives a value that greatly exceeds the nominal one, the proposed tests using all the data with type I error rates give a value close to the nominal level.
Second, the proposed test is more powerful than the usual test.

\section*{Acknowledgments}
This work was supported by JSPS KAKENHI (Grant Numbers 20K03756 and 20K11714).
\def\thesection{Appendix}
\def\thesubsection{A.\arabic{subsection}}
\section{}
\label{sec.Ap}
\def\theequation{A.\arabic{equation}}
\def\thethm{A.\arabic{thm}}
\def\thelem{A.\arabic{lem}}

This section provides the algorithms for the hypotheses $H_{S^{a}}$ and $H_{S^{b}}$.

\subsection*{(i) Hypothesis $H_{S^{a}}$}
The E-step and M-step under the hypothesis $H_{S^{a}}$ are obtained as described below.
\begin{flushleft}
\textbf{E-step}
\end{flushleft}
The $Q^*$-function is given as
\begin{align*}
    Q^*(\bm{\Theta}|\bm{\Theta}^{(t)}) 
    &=\sum_{i}\sum_{j}n_{00,ij}^{(t+1)}\log \pi_{00,ij}+n_{\alpha}^{(t+1)}\log \alpha \\
    & \quad +n_{\beta_{1}}^{(t+1)}\log \beta_1+n_{\beta_{2}}^{(t+1)}\log \beta_{2},
\end{align*}
where $\{n_{00,ij}^{(t+1)}\}$, $n_{\alpha}^{(t+1)}$, and $\{n_{\beta_{k}}^{(t+1)}\}$ are defined as
\begin{align*}
    n_{00,ij}^{(t+1)} &= n_{00,ij} + n_{01,i+}\cdot\frac{\beta_{j}^{(t)}\pi_{00,ij}^{(t)}}{\beta_{1}^{(t)}\pi_{00,i1}^{(t)}+\beta_{2}^{(t)}\pi_{00,i2}^{(t)}}\\
    & \quad +n_{10,+j}\cdot\frac{\pi_{00,ij}^{(t)}}{\pi_{00,\bullet j}^{(t)}}+n_{11,++}\cdot\frac{\beta_j^{(t)}\pi_{00,ij}^{(t)}}{\sum_{u} \beta_u^{(t)}\pi_{00,\bullet u}^{(t)}} \quad (i=1,2; j=1,2),\\
    n_{\alpha}^{(t+1)} &= n_{10,+1} + n_{10,+2} + n_{11,++}, \\
    n_{\beta_{k}}^{(t+1)} &= n_{01,1+}\cdot\frac{\beta_k^{(t)}\pi_{00,1k}^{(t)}}{\beta_1^{(t)}\pi_{00,11}^{(t)}+\beta_2^{(t)}\pi_{00,12}^{(t)}}+n_{01,2+}\cdot\frac{\beta_k^{(t)}\pi_{00,2k}^{(t)}}{\beta_1^{(t)}\pi_{00,21}^{(t)}+\beta_2^{(t)}\pi_{00,22}^{(t)}} \\
    & \quad +n_{11,++}\cdot\frac{\beta_k^{(t)}\pi_{00,\bullet k}^{(t)}}{\sum_{u} \beta_u^{(t)} \pi_{00,\bullet u}^{(t)}} \quad (k=1,2).\\
\end{align*}
Note that $n_{\alpha}^{(t+1)}$ is a constant.
\begin{flushleft}
\textbf{M-step}
\end{flushleft}
\begin{description}
    \item[1st step]\begin{align*}
        \alpha^{(t+1)} =\frac{n_{\alpha}^{(t+1)}}{\lambda_{1}^{(t)} A^{(t)} + \lambda_{2}^{(t)} B^{(t)} },
    \end{align*}
    where
    \begin{align*}
        A^{(t)} &= \left( 1 + \beta_{1}^{(t)} \right) \pi_{00,\bullet 1}^{(t)} + \left( 1 + \beta_{2}^{(t)} \right) \pi_{00,\bullet 2}^{(t)}, \\
        B^{(t)} &= \left( 1 + \beta_{2}^{(t)} \right) \pi_{00,12}^{(t)} - \left( 1 + \beta_{1}^{(t)} \right) \pi_{00,21}^{(t)}.
    \end{align*}
    \item[2nd step]\begin{align*}
        \beta_1^{(t+1)} &=\frac{n_{\beta_{1}}^{(t+1)}}{\left(1+\alpha^{(t+1)} \right)\left(\lambda_{1}^{(t)}\pi_{00,\bullet 1}^{(t)}-\lambda_{2}^{(t)}\pi_{00,21}^{(t)}\right)}, \\ 
        \beta_2^{(t+1)} &=\frac{n_{\beta_{2}}^{(t+1)}}{\left(1+\alpha^{(t+1)} \right)\left(\lambda_{1}^{(t)}\pi_{00,\bullet 2}^{(t)}+\lambda_{2}^{(t)}\pi_{00,12}^{(t)}\right)}.
    \end{align*}
    \item[3rd step]\begin{align*}
        \pi_{00,11}^{(t+1)} &=\frac{n_{00,11}^{(t+1)}}{\lambda_{1}^{(t)}\left(1+\alpha^{(t+1)}+\beta_{1}^{(t+1)}+\alpha^{(t+1)}\beta_{1}^{(t+1)}\right)},\\
        \pi_{00,12}^{(t+1)} &=\frac{n_{00,12}^{(t+1)}+n_{00,21}^{(t+1)}}{2\lambda_{1}^{(t)}\left(1+\alpha^{(t+1)}+\beta_{2}^{(t+1)}+\alpha^{(t+1)}\beta_{2}^{(t+1)}\right)},\\
        \pi_{00,21}^{(t+1)} &=\frac{n_{00,12}^{(t+1)}+n_{00,21}^{(t+1)}}{2\lambda_{1}^{(t)}\left(1+\alpha^{(t+1)}+\beta_{1}^{(t+1)}+\alpha^{(t+1)}\beta_{1}^{(t+1)}\right)},\\
        \pi_{00,22}^{(t+1)} &=\frac{n_{00,22}^{(t+1)}}{\lambda_{1}^{(t)}\left(1+\alpha^{(t+1)}+\beta_{2}^{(t+1)}+\alpha^{(t+1)}\beta_{2}^{(t+1)}\right)},\\
        \lambda_{1}^{(t+1)} &=n_{00,11}^{(t+1)}+n_{00,12}^{(t+1)}+n_{00,21}^{(t+1)}+n_{00,22}^{(t+1)},\\
        \lambda_{2}^{(t+1)} &=\frac{\lambda_{1}^{(t)}\left(n_{00,12}^{(t+1)}-n_{00,21}^{(t+1)}\right)}{n_{00,12}^{(t+1)}+n_{00,21}^{(t+1)}}.
        \end{align*}
\end{description}
Note that $\lambda_{1}^{(t+1)}$ is the sample size $n$.

\subsection*{(ii) Hypothesis $H_{S^{b}}$}
The E-step and M-step under the hypothesis $H_{S^{b}}$ are obtained as described below.
\begin{flushleft}
\textbf{E-step}
\end{flushleft}
The $Q^*$-function is given as
\begin{align*}
    Q^*(\bm{\Theta}|\bm{\Theta}^{(t)}) 
    &=\sum_{i}\sum_{j}n_{00,ij}^{(t+1)}\log \pi_{00,ij}+n_{\alpha_{1}}^{(t+1)}\log \alpha_1+n_{\alpha_{2}}^{(t+1)}\log \alpha_2 \\
    & \quad +n_{\beta_{1}}^{(t+1)}\log \beta_{1} + n_{\beta_{2}}^{(t+1)}\log \beta_{2},
\end{align*}
where $\{n_{00,ij}^{(t+1)}\}$, $\{n_{\alpha_{k}}^{(t+1)}\}$, and $\{n_{\beta_{k}}^{(t+1)}\}$ are defined as
\begin{align*}
    n_{00,ij}^{(t+1)} &= n_{00,ij} + n_{01,i+}\cdot\frac{\beta_{j}^{(t)}\pi_{00,ij}^{(t)}}{\beta_{1}^{(t)}\pi_{00,i1}^{(t)}+\beta_{2}^{(t)}\pi_{00,i2}^{(t)}}+n_{10,+j}\cdot\frac{\alpha_i^{(t)}\pi_{00,ij}^{(t)}}{\alpha_1^{(t)}\pi_{00,1j}^{(t)}+\alpha_2^{(t)}\pi_{00,2j}^{(t)}} \\
    & \quad +n_{11,++}\cdot\frac{\alpha_i^{(t)}\beta_j^{(t)}\pi_{00,ij}^{(t)}}{\sum_{s}\sum_{u} \alpha_s^{(t)}\beta_u^{(t)}\pi_{00,su}^{(t)}} \quad (i=1,2;j=1,2),\\
    n_{\alpha_{k}}^{(t+1)} &= n_{10,+1}\cdot\frac{\alpha_k^{(t)}\pi_{00,k1}^{(t)}}{\alpha_1^{(t)}\pi_{00,11}^{(t)}+\alpha_2^{(t)}\pi_{00,21}^{(t)}}+n_{10,+2}\cdot\frac{\alpha_k^{(t)}\pi_{00,k2}^{(t)}}{\alpha_1^{(t)}\pi_{00,12}^{(t)}+\alpha_2^{(t)}\pi_{00,22}^{(t)}} \\
    & \quad +n_{11,++}\cdot\frac{\alpha_k^{(t)} \left(\beta_{1}^{(t)}\pi_{00,k1}^{(t)} + \beta_{2}^{(t)}\pi_{00,k2}^{(t)} \right)}{\sum_{s}\sum_{u} \alpha_s^{(t)}\beta_u^{(t)} \pi_{00,su}^{(t)}} \quad (k=1,2),\\
    n_{\beta_{k}}^{(t+1)} &= n_{01,1+}\cdot\frac{\beta_k^{(t)}\pi_{00,1k}^{(t)}}{\beta_1^{(t)}\pi_{00,11}^{(t)}+\beta_2^{(t)}\pi_{00,12}^{(t)}}+n_{01,2+}\cdot\frac{\beta_k^{(t)}\pi_{00,2k}^{(t)}}{\beta_1^{(t)}\pi_{00,21}^{(t)}+\beta_2^{(t)}\pi_{00,22}^{(t)}} \\
    & \quad +n_{11,++}\cdot\frac{\beta_k^{(t)} \left( \alpha_{1}^{(t)}\pi_{00,1k}^{(t)} + \alpha_{2}^{(t)}\pi_{00,2k}^{(t)}\right)}{\sum_{s}\sum_{u} \alpha_{s}^{(t)}\beta_u^{(t)} \pi_{00,su}^{(t)}} \quad (k=1,2).\\  
\end{align*}
\begin{flushleft}
\textbf{M-step}
\end{flushleft}
\begin{description}
    \item[1st step]\begin{align*}
        \alpha_1^{(t+1)} &=\frac{n_{\alpha_{1}}^{(t+1)}}{\lambda_{1}^{(t)}\left( \left(\beta^{(t)}_{1}+1\right)\pi_{00,11}^{(t)} + \left(\beta^{(t)}_{2}+1\right)\pi_{00,12}^{(t)} \right) +\lambda_{2}^{(t)}\left(\beta^{(t)}_{2}+1\right)\pi_{00,12}^{(t)}}, \\ 
        \alpha_2^{(t+1)} &=\frac{n_{\alpha_{2}}^{(t+1)}}{\lambda_{1}^{(t)}\left( \left(\beta^{(t)}_{1}+1\right)\pi_{00,21}^{(t)} + \left(\beta^{(t)}_{2}+1\right)\pi_{00,22}^{(t)} \right) -\lambda_{2}^{(t)}\left(\beta^{(t)}_{1}+1\right)\pi_{00,21}^{(t)}}.
    \end{align*}
    \item[2nd step]\begin{align*}
        \beta_1^{(t+1)} &=\frac{n_{\beta_{1}}^{(t+1)}}{\lambda_{1}^{(t)}\left( \left(\alpha^{(t+1)}_{1}+1\right)\pi_{00,11}^{(t)} + \left(\alpha^{(t+1)}_{2}+1\right)\pi_{00,21}^{(t)} \right) -\lambda_{2}^{(t)}\left(\alpha^{(t+1)}_{2}+1\right)\pi_{00,21}^{(t)}}, \\ 
        \beta_2^{(t+1)} &=\frac{n_{\beta_{2}}^{(t+1)}}{\lambda_{1}^{(t)}\left( \left(\alpha^{(t+1)}_{1}+1\right)\pi_{00,12}^{(t)} + \left(\alpha^{(t+1)}_{2}+1\right)\pi_{00,22}^{(t)} \right) +\lambda_{2}^{(t)}\left(\alpha^{(t+1)}_{1}+1\right)\pi_{00,12}^{(t)}}.
    \end{align*}
    \item[3rd step]\begin{align*}
        \pi_{00,11}^{(t+1)} &=\frac{n_{00,11}^{(t+1)}}{\lambda_{1}^{(t)}\left(1+\alpha_{1}^{(t+1)}+\beta_{1}^{(t+1)}+\alpha_1^{(t+1)}\beta_{1}^{(t+1)}\right)},\\
        \pi_{00,12}^{(t+1)} &=\frac{n_{00,12}^{(t+1)}+n_{00,21}^{(t+1)}}{2\lambda_{1}^{(t)}\left(1+\alpha_{1}^{(t+1)}+\beta_{2}^{(t+1)}+\alpha_1^{(t+1)}\beta_{2}^{(t+1)}\right)},\\
        \pi_{00,21}^{(t+1)} &=\frac{n_{00,12}^{(t+1)}+n_{00,21}^{(t+1)}}{2\lambda_{1}^{(t)}\left(1+\alpha_{2}^{(t+1)}+\beta_{1}^{(t+1)}+\alpha_2^{(t+1)}\beta_{1}^{(t+1)}\right)},\\
        \pi_{00,22}^{(t+1)} &=\frac{n_{00,22}^{(t+1)}}{\lambda_{1}^{(t)}\left(1+\alpha_{2}^{(t+1)}+\beta_{2}^{(t+1)}+\alpha_2^{(t+1)}\beta_{2}^{(t+1)}\right)},\\
        \lambda_{1}^{(t+1)} &=n_{00,11}^{(t+1)}+n_{00,12}^{(t+1)}+n_{00,21}^{(t+1)}+n_{00,22}^{(t+1)},\\
        \lambda_{2}^{(t+1)} &=\frac{\lambda_{1}^{(t)}\left(n_{00,12}^{(t+1)}-n_{00,21}^{(t+1)}\right)}{n_{00,12}^{(t+1)}+n_{00,21}^{(t+1)}}.
        \end{align*}
\end{description}
Note that $\lambda_{1}^{(t+1)}$ is the sample size $n$.


\end{document}